\documentclass[12pt]{article}

\usepackage{epsfig,cite,amsmath,amsfonts,amssymb}
\usepackage{graphicx}

\allowdisplaybreaks \overfullrule=0pt \textheight= 23 truecm
\newcommand\mysection{\setcounter{equation}{0}\section}

\newcounter{hran} 
\def\ba{\begin{array}} \def\ea{\end{array}}
\def\lsim{\raise0.3ex\hbox{$<$\kern-0.75em\raise-1.1ex\hbox{$\sim$}}}
\def\gsim{\raise0.3ex\hbox{$>$\kern-0.75em\raise-1.1ex\hbox{$\sim$}}}
\def\to{\rightarrow} \def\ie{{\it i.e. }} \def\nn{\nonumber}
\def\noi{\noindent} \def\vev#1{\langle #1 \rangle} \def\gev{~{\rm
GeV}} \def\tev{~{\rm TeV}} \def\a{\alpha} \def\b{\beta} \def\d{\delta}
\def\D{\Delta}  \def\k{\kappa} \def\l{\lambda}
\def\p{\varphi} \def\t{\theta} \def\eK{\varepsilon_K}
\def\vckm{V_{\mathrm{CKM}}} \def\wino{\tilde{W}}
\def\higgsino{\tilde{H}} \def\st#1{\tilde{t}_{#1}}
\def\sb#1{\tilde{b}_{#1}} \def\charg{\tilde{\chi}^{\pm}}
\def\chargi#1{\tilde{\chi}^{\pm}_{#1}}

\begin{document}

\pagestyle{empty}
\begin{flushright}
FTUAM 03/06 \\
IFT-UAM/CSIC-03-12 \\
IFIC/03-08 \\ 
hep-ph/0304xxx
\end{flushright}
\vskip 2 truecm

\begin{center}
{\Large\bf Spontaneous CP Violation in \\ Non-Minimal Supersymmetric Models}
\end{center}
\vskip 2 truecm

\centerline{\bf Cyril Hugonie$^{\;a}$, Jorge C. Rom\~ao$^{\;b}$, Ana M.
Teixeira$^{\;b,\,c}$}
\vskip 5 truemm

\centerline{$^{a}$ AHEP Group, Instituto de F\'\i sica Corpuscular --
CSIC/Universitat de Val\`encia}
\vskip 1 truemm

\centerline{Edificio Institutos de Investigaci\'on, Apartado de Correos 22085,
E-46071 Val\`encia, Spain}
\vskip 3 truemm

\centerline{$^{b}$ Departamento de F\'\i sica and CFIF, Instituto Superior
T\'ecnico}
\vskip 1 truemm

\centerline{Av. Rovisco Pais, 1049-001 Lisboa, Portugal}
\vskip 3 truemm

\centerline{$^{c}$ Departamento de F\'\i sica Te\'orica C-XI and Instituto de
F\'\i sica Te\'orica C-XVI,} 
\vskip 1 truemm

\centerline{Universidad Aut\'onoma de Madrid, Cantoblanco, 28049 Madrid, Spain}

\vskip 1 truecm

\begin{abstract}
We study the possibilities of spontaneous CP violation in the Next-to-Minimal
Supersymmetric Standard Model with an extra singlet tadpole term in the scalar
potential. We calculate the Higgs boson masses and couplings with radiative
corrections including dominant two loop terms. We show that it is possible to
satisfy the LEP constraints on the Higgs boson spectrum with non-trivial
spontaneous CP violating phases. We also show that these phases could account
for the observed value of $\varepsilon_K$.
\end{abstract}

\newpage \pagestyle{plain}

\mysection{Introduction}

The understanding of CP violation, first observed in $K$ decays \cite{Kaon},
remains an open and most challenging question in particle physics. In the
Standard Model (SM), CP violation arises from the presence of complex Yukawa
couplings in the lagrangian. In electroweak interactions, CP violation
originates in the misalignment of mass and charged-current interaction
eigenstates, and is parametrized by the physical phase of the
Cabibbo-Kobayashi-Maskawa (CKM) quark mixing matrix \cite{KM}, $\delta_{CKM}$.
Although SM predictions are in good agreement with experimental observations,
including recent measurements at $B$-factories, the SM amount of CP violation
fails to account for the observed baryon asymmetry of the Universe \cite{BAU}. 
Moreover, one is yet to find an answer for the strong CP problem, or in other
words, to understand the smallness of the $\bar{\theta}$ parameter. Bounds from
the electric dipole moment (EDM) of the neutron force this flavour-conserving
CP violating phase to be as small as $10^{-10}$, and this fine-tuning  is most
unnatural in the sense of 't~Hooft \cite{Hooft}, since the lagrangian does not
acquire any new symmetry in the limit where $\bar{\theta}$ vanishes.

In supersymmetric (SUSY) extensions of the SM there are additional sources of
explicit CP violation, arising from complex soft SUSY breaking terms  as well
as from the complex SUSY conserving $\mu$ parameter. As pointed in
\cite{susycp}, these phases can account for the values of the $K$ and $B$ meson
CP violating observables, even in the absence of $\delta_{CKM}$. However, the
supersymmetric phases also generate large contributions to the EDMs of the
electron, neutron and mercury atom. The non-observation of the EDMs imposes
strong constraints on the SUSY phases, forcing them to be very small. Putting
these new phases to zero is also not natural in the sense of 't Hooft. This is
the so-called SUSY CP problem and many are the solutions that have been
proposed to overcome it (for a review see Ref.~\cite{Khalil}).

An attractive approach to the SUSY CP problem is to impose CP invariance on the
lagrangian, and spontaneously break it through complex vacuum expectation
values (VEVs) for the Higgs scalar fields \cite{LW}. Thus, CP symmetry is
restored at high energy and CP violating phases appear as dynamical variables. 
Spontaneous CP violation (SCPV) is also an appealing solution to the strong CP
problem, since in this case one naturally has a vanishing $\bar{\theta}$ at
tree level~\cite{strongcp}.  Further motivation to SCPV stems from string
theories, where it has been shown that in string perturbation theory CP exists
as a good symmetry that could be spontaneously broken~\cite{StrWit}.

SCPV requires at least two Higgs doublets. The Minimal Supersymmetric Standard
Model (MSSM) is a very appealing example of a two Higgs doublet model. However,
it is well known that, at tree level, SCPV does not occur in the MSSM
\cite{HHG}. On the other hand, radiative corrections can generate CP violating
operators \cite{Maekawa} but then, according to the Georgi-Pais theorem on
radiatively broken global symmetries \cite{GP}, one expects to have light Higgs
states in the spectrum \cite{Pomarol}, which are excluded by LEP
\cite{LEP,LEPHA,LEPCH}.

The case of the Next-to-Minimal Supersymmetric Standard Model (NMSSM)
\cite{NMSSM,PRD39}, where a singlet superfield is added to the Higgs sector, is
more involved. In the usual $\mathbb{Z}_3$ invariant version of the model,
where only dimensionless couplings are allowed in the superpotential, it has
been shown that, although CP violating extrema are present at tree level, these
are always maxima, not minima of the scalar potential, with negative squared
masses for the Higgs states \cite{Jorge}. Hence SCPV is not feasible at tree
level in the NMSSM. Furthermore, as in the MSSM case, radiatively induced CP
violating minima always bear light Higgs states which are difficult to
accommodate with LEP data \cite{BB}. The possibility of SCPV has been studied in
more general non-minimal models, where no $\mathbb{Z}_3$ symmetry is assumed
and dimensionful, SUSY conserving terms are present in the superpotential
\cite{Pomarol2,Glasgow,Lisbon}. In this case, it has been shown that SCPV is
possible and could account for the observed value of $\eK$ in the kaon system
\cite{Pomarol2,Lisbon}. 

The first drawback of the general NMSSM with respect to the $\mathbb{Z}_3$
invariant version, is that it no longer provides a solution to the $\mu$
problem of the MSSM, which was one of the original motivations of the NMSSM.
The second is that, in the absence of a global symmetry under which the singlet
field is charged, divergent singlet tadpoles proportional to
$M_{\mathrm{Planck}}$, generated by non-renormalizable higher order
interactions, can appear in the effective scalar potential
\cite{Ulrich,tadpoles}. Such tadpole terms would destabilize the hierarchy
between the electroweak (EW) scale and the Planck scale. On the other hand, if
$\mathbb{Z}_3$, or any other discrete symmetry, is imposed at the lagrangian
level, it is spontaneously broken at the EW scale once the Higgs fields get
non-vanishing VEVs, giving rise to disastrous cosmological domain walls
\cite{walls}. It has recently been argued that using global discrete
$R$-symmetries for the complete theory - including non-renormalizable
interactions - one could construct a $\mathbb{Z}_3$ invariant renormalizable
superpotential and generate a $\mathbb{Z}_3$ breaking non-divergent singlet
tadpole term in the scalar potential \cite{Tamvakis,MNSSM}. These models are
free of both stability and domain wall problems, and all the dimensionful
parameters, including the singlet tadpole, are generated through the soft SUSY
breaking terms.

The aim of this paper is to study the possibility of SCPV in the NMSSM with an
extra singlet tadpole term in the effective potential, taking into account the
latest experimental constraints on Higgs boson and sparticle masses, as well as
the observed value of $\eK$. In particular, contrary to what was asserted in
Ref.~\cite{Habaetal}, we obtain that in the $\mathbb{Z}_3$ symmetric limit,
SCPV {\it cannot} be accommodated with the LEP exclusion limit on a light Higgs
boson. On the other hand, we show that if $\mathbb{Z}_3$ is broken by a
non-zero singlet tadpole, it is possible to spontaneously break CP, satisfy the
LEP constraints on the Higgs boson mass and have $\eK$ compatible with the
experimental value.  The paper is organized as follows: in section~2 we define
the model and derive the Higgs boson mass matrix. The procedure to scan the
parameter space and the resulting Higgs boson spectrum are discussed in
section~3. Section~4 is devoted to the calculation of $\eK$. Finally, we
present the conclusions in section~5.

\mysection{Overview of the model}

\subsection{Higgs scalar potential}

In addition to the Yukawa couplings for the quarks and leptons (as in the
MSSM), the superpotential of the NMSSM is defined by

\begin{equation} \label{supot} 
W_{\mathrm{Higgs}} = \l \hat{H}_1 \hat{H}_2 \hat{N} + \frac{1}{3} \k \hat{N}^3
\; ,
\end{equation}

\noi where $\hat{H}_1 = (\hat{H}_1^0, \hat{H}_1^-)$ is the Higgs doublet
superfield coupled to the down-type fermions, $\hat{H}_2 = (\hat{H}_2^+,
\hat{H}_2^0)$ the one coupled to the up-type ones, and $\hat{N}$ is a singlet.
Once EW symmetry is broken, the scalar component of $\hat{N}$ acquires a VEV,
$x = |\vev N|$, thus generating an effective $\mu$ term

\begin{equation} \label{defmu}
\mu \equiv \lambda x \; .
\end{equation}

The superpotential in Eq.~(\ref{supot}) is scale invariant, and the EW scale
appears only through the soft SUSY breaking terms. It is also invariant under a
global $\mathbb{Z}_3$ symmetry. The possible domain wall problem due to the
spontaneous breaking of the $\mathbb{Z}_3$ symmetry at the EW scale is assumed
to be solved by adding non-renormalizable interactions which break the
$\mathbb{Z}_3$ symmetry without spoiling the quantum stability with unwanted
divergent singlet tadpoles. This can be achieved by replacing the
$\mathbb{Z}_3$ symmetry by a set of discrete $R$-symmetries, broken by the soft
SUSY breaking terms \cite{Tamvakis,NMSSM}. At low energy,  the additional
non-renormalizable terms allowed by the $R$-symmetries generate an extra linear
term for the singlet in the effective potential, through tadpole loop diagrams

\begin{equation} \label{Vtad}
V_{\mathrm{tadpole}} = - \xi^3 N + \mathrm{h.c.} \; ,
\end{equation}

\noi where $\xi$ is of the order of the soft SUSY breaking terms ($\lsim$ 1
TeV). Since our approach is phenomenological, we take $\xi$ as a free
parameter, without considering the details of the non-renormalizable
interactions that generate it. Likewise, we do not discard the singlet
self-coupling term in Eq.~(\ref{supot}) by imposing $\k = 0$, which is possible
once $\xi \neq 0$ \cite{MNSSM}, but rather assume $\k$ to be a free parameter.

In addition to $V_{\mathrm{tadpole}}$, the tree level Higgs potential has the
usual $F$ and $D$ terms as well as soft SUSY breaking terms:

\begin{eqnarray} \label{treepot} 
V_F & = & \l^2 \left( |H_1|^2 |N|^2 + |H_2|^2 |N|^2 + |H_1|^2 |H_2|^2 \right)
\nn \\
& & - \l^2 \left( H_1^{0*}H_2^{0*}H_1^-H_2^+ + \mathrm{h.c.} \right) + \k^2
|N|^4 \; , \nn \\
V_D & = & \frac{g_1^2+g_2^2}{8} \left( |H_1|^2 - |H_2|^2 \right)^2 +
\frac{g_2^2}{2} |H_1^\dagger H_2|^2 \; , \\
V_\mathrm{soft} & = & m_{H_1}^2 |H_1|^2 + m_{H_2}^2 |H_2|^2 + m_N^2 |N|^2 \nn
\\
& & + \left( \l A_\l N H_1 H_2 + \frac{1}{3} \k A_\k N^3 + \mathrm{h.c.}
\right) \; , \nn
\end{eqnarray}

\noi where $g_1$ and $g_2$ are the $U(1)_Y$ and $SU(2)_L$ coupling constants,
respectively. In what follows, the soft SUSY breaking terms $m_{H_1}, m_{H_2},
m_N, A_\l, A_\k$ are taken as free parameters at the weak scale and no
assumption is made on their value at the GUT scale. We assume that the
lagrangian is CP invariant, which means that all the parameters appearing in
Eqs.~(\ref{Vtad}, \ref{treepot}) are real. On the other hand, once the EW
symmetry is spontaneously broken, the neutral Higgs fields acquire complex VEVs
that spontaneously break CP. By gauge invariance, one can take $\vev{H_1^-} =
0$. The condition for a local minimum with $\vev{H_2^+} =0$ is equivalent to a
positive square mass for the charged Higgs boson. The VEVs of the neutral Higgs
fields have the general form

\begin{equation} 
\vev{H_1^0} = v_1 e^{i\p_1} \; , \quad \vev{H_2^0} = v_2 e^{i\p_2} \; , \quad
\vev{N} = x e^{i\p_3} \; ,
\end{equation}

\noi where $v_1, v_2, x$ are positive and $\varphi_1, \varphi_2, \varphi_3$
are CP violating phases. However, only two of these phases are physical. They
can be chosen as

\begin{equation} 
\t = \p_1+\p_2+\p_3 \quad \mathrm{and} \quad \d = 3 \p_3 \; .
\end{equation}

\subsection{Minimization of the tree level potential}

From the tree level scalar potential in Eqs.~(\ref{Vtad}, \ref{treepot}), one
can derive the five minimization equations for the VEVs and phases $v_1, v_2,
x, \t, \d$. They can be used to express the soft parameters $m_{H_1}, m_{H_2}$,
$m_N, A_\l, A_\k$ in terms of $v_1, v_2, x, \t, \d$:

\begin{eqnarray} \label{mineqtree}
\frac{\partial V_\mathrm{tree}}{\partial v_1} = 0 \quad \Rightarrow \quad
m_{H_1}^2 & = & -\l^2 \left( x^2 + v^2\sin^2\b \right) - \frac{1}{2} M_Z^2 \cos
2\b \nn \\
& & - \l x \tan\b \left( \k x \cos(\t-\d) + A_\l \cos\t \right) \; , \nn \\
\frac{\partial V_\mathrm{tree}}{\partial v_2} = 0 \quad \Rightarrow \quad
m_{H_2}^2 & = & -\l^2 \left( x^2 + v^2\cos^2\b \right) + \frac{1}{2} M_Z^2 \cos
2\b \nn \\
& & - \l x \cot\b \left( \k x \cos(\t-\d) + A_\l \cos\t \right) \; , \nn \\
\frac{\partial V_\mathrm{tree}}{\partial x} = 0 \quad \Rightarrow \quad
m_{N}^2 & = & -\l^2 v^2 - 2\k^2 x^2 - \l\k v^2 \sin 2\b \cos(\t-\d) \\
& & - \frac{\l A_\l v^2}{2x} \sin 2\b \cos\t - \k A_\k x \cos\d +
\frac{\xi^3}{x} \cos(\d/3) \; , \nn \\
\frac{\partial V_\mathrm{tree}}{\partial \t} = 0 \quad \Rightarrow \quad
A_\l & = & - \frac{\k x \sin(\t-\d)}{\sin\t} \; , \nn \\
\frac{\partial V_\mathrm{tree}}{\partial \d} = 0 \quad \Rightarrow \quad
A_\k & = & \frac{3 \l\k x v^2 \sin 2\b \sin(\t-\d) + 2 \xi^3 \sin(\d/3)}{2 \k
x^2 \sin\d} \; , \nn
\end{eqnarray}

\noi with $\tan\b = v_2 / v_1$, $v = \sqrt{v_1^2+v_2^2} = 174\gev$ and $M_Z$
the $Z$ boson mass. The above relations allow us to use $\tan\b, x, \t$ and
$\d$ instead of $m_{H_1}, m_{H_2}, m_N, A_\l, A_\k$ as free parameters.

Once EW symmetry is spontaneously broken, we are left with five neutral Higgs
states and a pair of charged Higgs states. The neutral Higgs fields can be
rewritten in terms of CP eigenstates

\begin{eqnarray}
H_1^0 & = & e^{i \p_1} \left \{ v_1 + \frac{1}{\sqrt{2}} (S_1 + i \sin\b P)
\right \} \; , \nn \\
H_2^0 & = & e^{i \p_2} \left \{ v_2 + \frac{1}{\sqrt{2}} (S_2 + i \cos\b P)
\right \} \; , \\
N & = & e^{i \p_3} \left \{ x + \frac{1}{\sqrt{2}} (X + i Y) \right \} \; , \nn
\end{eqnarray}

\noi where $S_1, S_2, X$ are the CP-even components, and $P, Y$ are the CP-odd
components. Note that we have rotated away the CP-odd would-be Goldstone boson
associated with the EW symmetry breaking. The mass matrix for the neutral Higgs
bosons in the basis $(S_1, S_2, X, P, Y)$ can be easily obtained:

\begin{eqnarray} \label{mamatree}
M_{11}^2 & = & M_Z^2 \cos^2\b - \frac{\l\k x^2 \tan\b \sin\d}{\sin\t} \nn \\
M_{22}^2 & = & M_Z^2 \sin^2\b - \frac{\l\k x^2 \cot\b \sin\d}{\sin\t} \nn \\
M_{33}^2 & = & 4 \k^2 x^2 + \frac{1}{2} \l\k v^2 \sin2\b\sin(\t-\d) \left(
\cot\t + 3\cot\d \right) + \frac{\xi^3}{x} \left( \sin(\d/3)\cot\d + \cos(\d/3)
\right) \nn \\
M_{44}^2 & = & - \frac{2 \l\k x^2 \sin\d}{\sin 2\b \sin\t} \nn \\
M_{55}^2 & = & \frac{1}{2} \l\k v^2 \sin2\b \left( 6 \cos(\t-\d) - 9
\frac{\sin\t}{\sin\d} - \frac{\sin\d}{\sin\t} \right) + \frac{\xi^3}{x} \left(
\cos(\d/3) - 3 \sin(\d/3)\cot\d \right) \nn \\
M_{12}^2 & = & \left( \l^2 v^2 - \frac{M_Z^2}{2} \right) \sin 2\b + \frac{\l\k
x^2 \sin\d}{\sin\t} \nn \\
M_{13}^2 & = & 2 \l^2 x v \cos\b + \l\k x v \sin\b \left( \cos(\t-\d) +
\frac{\sin\d}{\sin\t} \right) \\
M_{14}^2 & = & 0 \nn \\
M_{15}^2 & = & 3 \l\k x v \sin\b \sin(\t-\d) \nn \\
M_{23}^2 & = & 2 \l^2 x v \sin\b + \l\k x v \cos\b \left( \cos(\t-\d) +
\frac{\sin\d}{\sin\t} \right) \nn \\ 
M_{24}^2 & = & 0 \nn \\
M_{25}^2 & = & 3 \l\k x v \cos\b \sin(\t-\d) \nn \\
M_{34}^2 &= & - \l\k x v \sin(\t-\d) \nn \\
M_{35}^2 & = & -2 \l\k v^2 \sin 2\b \sin(\t-\d) \nn \\
M_{45}^2 & = & \l\k x v \left( \cos(\t-\d) + \frac{\sin\d}{\sin\t} \right) \nn 
\end{eqnarray}

\noi where we made use of the minimization conditions in Eq.~(\ref{mineqtree})
to eliminate the soft terms and simplify the expressions. One can note that
$M_{14}^2=M_{24}^2=0$, which means that there is {\it no CP violation in the
Higgs doublet sector}. On the other hand, CP violating mixings between the
singlet and the doublets can appear, as long as $\t \neq \d$. It is easy to see
from Eq.~(\ref{mamatree}) that, if $\xi \neq 0$, one can evade the NMSSM no-go
theorem for SCPV: it was shown in \cite{Jorge} that the tree level mass matrix
for the neutral Higgs bosons always had one negative eigenvalue in the case of
non-trivial phases of the VEVs. However, in our case the negative eigenvalue
can be lifted up to a positive value provided $\xi$ is large enough, due to the
additional diagonal terms proportional to $\xi^3$ in $M_{33}^2$ and $M_{55}^2$.
Hence, SCPV is possible already {\it at tree level} for $\xi \neq 0$. We will
check this numerically in section \ref{sec:higgs}, taking into account
radiative corrections as well as experimental constraints on the Higgs boson
masses.

\subsection{Radiative corrections}

It is well known that one loop radiative corrections can give large
contributions to the Higgs boson masses in the MSSM \cite{MSSM1loop} as well as
in the NMSSM \cite{NMSSM1loop}. Furthermore, they play a crucial role in the
SCPV mechanism \cite{Maekawa,BB}, as they generate CP violating operators. In
what follows, we shall only consider radiative corrections due to top-stop
loops. The one loop effective potential reads

\begin{equation} \label{1loopot} 
V_{\mathrm{1-loop}} = \frac{3}{32\pi^2} \left\{ {\cal M}_{\st{i}}^4 \left (
\log \frac{{\cal M}_{\st{i}}^2}{Q^2} - \frac{3}{2} \right ) - 2 {\cal M}_t^4
\left ( \log \frac{{\cal M}_t^2}{Q^2} - \frac{3}{2} \right ) \right\} \; .
\end{equation}

\noi ${\cal M}_t^2$ and ${\cal M}_{\st{i}}^2, (i=1,2)$ are the field dependent
top and stop squared masses respectively, and $Q$ is the renormalization scale,
at which all the parameters are evaluated. $Q$ has to be of the order of the
soft SUSY breaking parameters so that the tree level scalar potential has the
supersymmetric form of Eq.~(\ref{treepot}). The field dependent stop mass
matrix, in the basis $(T_L , T_R^{c*})$, is given by

\begin{equation} 
{\cal M}_{\tilde{t}}^2 = \left[ \ba{cc} m_Q^2 + h_t^2|H_2^0|^2 &
h_t(A_t H_2^{0*} + \l H_1^0 N) \\ h_t(A_t H_2^0 + \l H_1^{0*} N^*) &
m_T^2 + h_t^2|H_2^0|^2 \ea \right] \; ,
\end{equation}

\noi with $h_t$ the top Yukawa coupling and $m_Q, m_T, A_t$ the soft terms for
the stop sector. In the following, we assume $m_Q = m_T \equiv M_{\rm SUSY}$,
as this choice maximizes the radiative corrections to the lightest Higgs boson
mass. $D$ terms are not taken into account since we do not consider radiative
corrections proportional to the gauge couplings. At the minimum of the
potential, the top-stop masses are given by

\begin{eqnarray}
m_t^2 & = & h_t^2 v^2 \sin^2\b \; , \nn \\
m_{\st{1},\st{2}}^2 & = & M_{\rm SUSY} + m_t^2 \mp m_t X_t \; ,
\end{eqnarray}

\noi where

\begin{equation} \label{Xtdef}
X_t = \sqrt{A_t^2 + \l^2 x^2 \cot^2\b + 2 A_t \l x \cot\b \cos\t}
\end{equation}

\noi is the usual stop mixing parameter. This is not different from the CP
conserving case, up to the phase $\t$ appearing in $X_t$. One can show that
radiative corrections to the Higgs boson masses are minimized for $X_t = 0$
(minimal mixing scenario) and maximized for $X_t = \sqrt{6} M_{\rm SUSY}$
(maximal mixing scenario). However, in our case it is not possible to take $X_t
= 0$. In fact, the minimum of $X_t$ is $X_t = \l x \cot\b |\sin\t|$ for $A_t =
- \l x \cot\b \cos\t$. On the other hand, $X_t = \sqrt{6} M_{\rm SUSY}$ is only
possible if $\l x \cot\b |\sin\t| < \sqrt{6} M_{\rm SUSY}$. If this is not the
case, then the maximal mixing scenario is obtained for $X_t = \l x \cot\b
|\sin\t|$ and $A_t = - \l x \cot\b \cos\t$ as in the minimum mixing case.

The one loop terms give additional contributions to the minimization conditions
of Eq.~(\ref{mineqtree}). In particular, since $m_{\st{i}}^2$ depends on the CP
violating phase $\t$, the relation that gives $A_\l$ as a function of the VEVs
and phases of the Higgs fields is no longer valid. Moreover, as noted before,
the tree level minimization equations were used to simplify the Higgs boson mass
matrix elements at tree level. Nevertheless, it is possible to keep the tree
level mass matrix elements as in Eq.~(\ref{mamatree}) and write all the one
loop contributions as additional terms in the Higgs boson mass matrix

\begin{equation}
M_{ij}^2 \to M_{ij}^2 + \frac{3 h_t^2 m_t^2}{8\pi^2} \; \d M_{ij}^2 \; .
\end{equation}

One then obtains

\begin{eqnarray} \label{mamaloop}
\d M_{11}^2 & = & - \l^2 x^2 \D_1^2 \; f ( m_{\st{1}} , m_{\st{2}} ) \nn \\
\d M_{22}^2 & = & - A_t^2 \D_2^2 \; f ( m_{\st{1}} , m_{\st{2}} ) + 2 A_t \D_2
\; g ( m_{\st{1}} , m_{\st{2}} ) + \log \frac {m_{\st{1}}^2m_{\st{2}}^2}{m_t^4}
\nn \\
\d M_{33}^2 & = & - \l^2 v^2 \cos^2\b \D_1^2 \; f ( m_{\st{1}} , m_{\st{2}} )
\nn \\
\d M_{44}^2 & = & - \frac {\l^2 x^2 A_t^2 \sin^2\t}{\sin^2\b} \; f ( m_{\st{1}}
, m_{\st{2}} ) \nn \\
\d M_{55}^2 & = & - \l^2 v^2 \cos^2\b A_t^2 \sin^2\t \; f ( m_{\st{1}} ,
m_{\st{2}} ) \nn \\
\d M_{12}^2 & = & - \l x A_t \D_1 \D_2 \; f ( m_{\st{1}} , m_{\st{2}}) + \l x
\D_1 \; g ( m_{\st{1}} , m_{\st{2}} ) \nn \\
\d M_{13}^2 & = & - \l^2 x v \cos\b \D_1^2 \; f ( m_{\st{1}} , m_{\st{2}} ) +
\frac{\l^2 x v \cos\b}{m_t^2} \; h ( m_{\st{1}} , m_{\st{2}} ) \\
\d M_{14}^2 & = & \frac{\l^2 x^2 A_t \sin\t \D_1}{\sin\b} \; f ( m_{\st{1}} ,
m_{\st{2}} ) \nn \\
\d M_{15}^2 & = & \l^2 x v \cos\b A_t \sin\t \D_1 \; f ( m_{\st{1}} ,
m_{\st{2}} ) \nn \\
\d M_{23}^2 & = & - \l v \cos\b A_t \D_1 \D_2 \; f ( m_{\st{1}} , m_{\st{2}} )
+ \l v \cos\b \D_1 \; g ( m_{\st{1}} , m_{\st{2}} ) \nn \\
\d M_{24}^2 & = & \frac{\l x A_t^2 \sin\t \D_2}{\sin\b} \; f ( m_{\st{1}} ,
m_{\st{2}} ) - \frac{\l x A_t \sin\t}{\sin\b} \; g ( m_{\st{1}} , m_{\st{2}} ) 
\nn\\
\d M_{25}^2 & = & \l v \cos\b A_t^2 \sin\t \D_2 \; f ( m_{\st{1}} , m_{\st{2}}
) - \l v \cos\b A_t \sin\t \; g ( m_{\st{1}} , m_{\st{2}} ) \nn \\
\d M_{34}^2 & = & \frac{\l^2 x v \cos\b A_t \sin\t \D_1}{\sin\b} \; f (
m_{\st1} , m_{\st2} ) \nn \\
\d M_{35}^2 & = & \l^2 v^2 \cos^2\b A_t \sin\t \D_1 \; f ( m_{\st{1}} ,
m_{\st{2}} ) \nn \\
\d M_{45}^2 & = & - \frac{\l^2 x v \cos\b A_t^2 \sin^2\t}{\sin\b} \; f (
m_{\st{1}} , m_{\st{2}} ) \nn
\end{eqnarray}

\noi where

\begin{eqnarray} \label{loopfcn}
f(m_1,m_2) & = & \frac {1}{(m_2^2-m_1^2)^2} \left( \frac
{m_1^2+m_2^2}{m_2^2-m_1^2} \; \log \frac {m_2^2}{m_1^2} - 2 \right) \; , \nn \\
g(m_1,m_2) & = & \frac {1}{m_2^2-m_1^2} \; \log \frac {m_2^2}{m_1^2} \; , \\
h ( m_1 , m_2 ) & = & \frac {1}{m_2^2-m_1^2} \left [ m_2^2 \; \log \frac
{m_2^2}{Q^2} - m_1^2 \;\log \frac {m_1^2}{Q^2} \right ] - 1 \; , \nn
\end{eqnarray}

\noi and

\begin{equation}
\D_1 = A_t \cos\t + \l x \cot\b \; , \qquad \D_2 = A_t \cos\t + \l x \cot\b
\cos\t \; .
\end{equation}

\noi One can notice that, apart from a sub-leading term in $\d M_{13}^2$, all
the scale dependence is hidden in the parameters scale dependence. The one loop
terms of the neutral Higgs boson mass matrix in Eq.(\ref{mamaloop}) differ from
those given in Ref.~\cite{Habaetal}, where one loop minimization conditions
were not used to simplify the expressions.

Two loop corrections can also give substantial contributions to the Higgs boson
masses \cite{2loop,CH}. Our analysis follows closely the results of \cite{CH},
to which we refer the reader for more details. Here, we consider the dominant
two loop corrections which are proportional to $\a_s h_t^4$ and $h_t^6$, taking
only the leading logarithms (LLs) into account. In this approximation, the two
loop effective potential reads

\begin{equation} \label{2loopot}
V^{\mathrm{LL}}_{\mathrm{2-loops}} = 3 \left ( \frac{h_t^2}{16\pi^2} \right )^2
\left ( 32\pi\alpha_s - \frac{3}{2} \, h_t^2 \right ) v^4 \sin^4\b \; \log^2
\frac {m_t^2}{Q^2} \; .
\end{equation}

One loop corrections to the tree level relations between bare parameters and
physical observables, once reinserted in the one loop potential of
Eq.~(\ref{1loopot}), also appear as two loop effects. The dominant
contributions are:

\noi (\i) corrections to the kinetic terms of the Higgs bosons, which lead to
a wave function renormalization factor $Z_{H_2^0}$ given by

\begin{equation}
Z_{H_2^0} = 1 - 3 \frac{h_t^2}{16\pi^2} \; \log \frac {m_t^2}{Q^2} \; .
\end{equation}

\noi (\i\i) Corrections to the top quark Yukawa coupling

\begin{equation}
h_t(m_t) = h_t(Q) \left ( 1 + \frac{1}{32\pi^2} \left ( \frac{9}{2} \, h_t^2 -
32\pi\a_s \right ) \; \log \frac {m_t^2}{Q^2} \right ) \; .
\end{equation}

\noi The top quark running mass is then given by $m_t(m_t) = h_t(m_t)
Z_{H_2^0}^{1/2} v \sin\b$ and the relation between the pole and running masses
is, up to order $\alpha_s$, $m_t^{\rm pole} = (1 + \frac{4\alpha_s}{3\pi})
m_t(m_t)$. Similarly, the relation between the stop pole and running masses is
given by $m_{\st{i}}^{\rm pole} = (1 + \frac{8\alpha_s}{3\pi})
m_{\st{i}}(m_{\st{i}})$.

Once all these contributions are taken into account, one obtains a rather
complicated $5 \times 5$ mass matrix for the neutral Higgs fields, which can
only be numerically diagonalized (cf. section \ref{sec:higgs}).

\subsection{Charged Higgs boson mass}

Finally, we give the expression for the charged Higgs boson mass, which at tree
level is given by

\begin{equation}
m_{H^\pm}^2 = M_W^2 - \l^2 v^2 - \frac {2 \l\k x^2 \sin\d}{\sin2\b \sin\t} \; .
\end{equation}

\noi The radiative corrections due to top-bottom and stop-sbottom loops read

\begin{equation} \label{chrad}
\d m_{H^\pm}^2 = - \frac {3 h_t^2}{8 \pi^2} \frac {\l x A_t \cos\t}{\sin2\b} \;
h ( m_{\st{1}} , m_{\st{2}} ) + \d_r \; ,
\end{equation}

\noi where $h(m_1,m_2)$ has been defined in Eq.~(\ref{loopfcn}) and $\d_r$
contains the scale independent terms

\begin{eqnarray}
\d_r & = & \frac {3 h_t^2 m_t^2}{32 \pi^2} \frac {\l^2 x^2}{\sin^2\b} \left [
\frac {m_{\sb{L}}^2}{(m_{\st{1}}^2-m_{\sb{L}}^2) (m_{\st{2}}^2-m_{\sb{L}}^2)}
\; \log \frac {m_{\st{1}}^2 m_{\st{2}}^2}{m_{\sb{L}}^4} \right . \nn \\
& & \left . - \frac {1}{m_{\st{2}}^2-m_{\st{1}}^2} \left ( \frac
{m_{\st{1}}^2}{m_{\st{1}}^2-m_{\sb{L}}^2} + \frac
{m_{\st{2}}^2}{m_{\st{2}}^2-m_{\sb{L}}^2} \right ) \; \log \frac
{m_{\st{2}}^2}{m_{\st{1}}^2} \right ]\nn \\
& \simeq & - \frac {3 h_t^2}{16 \pi^2} \frac {\l^2 x^2 v^2}
{m_{\st{1}}^2+m_{\st{2}}^2} \; .
\end{eqnarray}

\noi In the above, we have assumed that the bottom Yukawa coupling is small
enough to be neglected, as justified by our choice of parameters, namely a low
$\tan\b$ regime. Therefore bottom squarks $\sb{L}$ and $\sb{R}$ do not mix and
$m_{\sb{L}} = m_Q \equiv M_{\rm SUSY}$. It is interesting to note that the term
proportional to $h ( m_{\st{1}} , m_{\st{2}} )$ in Eq.~(\ref{chrad}) exactly
cancels the one loop correction to $A_\l$ due to the minimization equation of
$V_{\mathrm{1-loop}}$ as a function of $\t$. Hence, if one replaces $A_\l$ by a
function of the VEVs and phases of the Higgs fields, one can simply rewrite the
charged Higgs boson mass as

\begin{equation} \label{chhmass}
m_{H^\pm}^2 \simeq M_W^2 - \l^2 v^2 - \frac {2 \l\k x^2 \sin\d}{\sin2\b \sin\t}
- \frac {3 h_t^2}{16 \pi^2} \frac {\l^2 x^2 v^2} {m_{\st{1}}^2+m_{\st{2}}^2} \; 
.
\end{equation}

\mysection{Mass spectrum} \label{sec:higgs}

In this section we investigate whether it is possible to have SCPV in the NMSSM
with the extra tadpole term for the singlet, given the exclusion limits on the
Higgs boson spectrum from LEP \cite{LEP,LEPHA,LEPCH}. In order to do so, we
perform a numerical scanning of the parameter space of the model. The
parameters appearing in the tree level Higgs boson mass matrix are $\l, \k,
A_\l, A_k, m_{H_1}, m_{H_2}, m_N, \xi$. As seen in the previous section, we can
use the minimization conditions of Eq.~(\ref{mineqtree}) to replace the soft
SUSY breaking terms by the VEVs and phases of the Higgs fields. In the
following, we will use the effective $\mu$ term defined in Eq.~(\ref{defmu}) as
a free parameter instead of the singlet VEV $x$, so that the free parameters of
the tree level Higgs boson mass matrix are now given by

\begin{equation} \label{freepar}
\l \; , \; \k \; , \; \tan\b \; , \; \mu \; , \; \xi \; , \; \t \; , \; \d \; .
\end{equation}

\noi Requiring the absence of Landau singularities for $\l$ and $\k$ below the
GUT scale ($M_{\mathrm GUT} \sim 2 \times 10^{16} \gev$) 
imposes upper bounds on
these couplings at the weak scale, which depend on the value of $h_t$, \ie of
$\tan\b$ and the top mass \cite{PRD39}. For $\tan\b \leq 10$ and $m_t^{\rm
pole} = 175 \gev$, one finds $\l_{\rm max} \sim 0.65$ and $\k_{\rm max} \sim
0.6$. This also yields a lower bound for $\tan\b$, namely $\tan\b \gsim 2.2$.
Regarding the radiative corrections, we take $M_{\rm SUSY} = 350 \gev$ and
assume the maximal mixing scenario for the stops, \ie $X_t = \max ( \sqrt{6}
M_{\rm SUSY}, \mu \cot\b |\sin\t| )$. The relatively small value for $M_{\rm
SUSY}$ will be justified in the following section when we address the
computation of $\eK$.

We have performed a numerical scanning on the free parameters, which were
randomly chosen in the following intervals:

\begin{eqnarray} \label{param}
& & 0.01 < \l < 0.65 \; , \qquad 0.01 < \k < 0.6 \; , \qquad 2.2 < \tan\b < 10
\; , \nn \\
& & 100 \gev < \mu < 500 \gev \; , \qquad 0 < \xi < 1 \tev \; , \\
& & -\pi < \t < \pi \; , \qquad -\pi < \d < \pi \nn \; .
\end{eqnarray}

\noi For each point, we computed the Higgs boson masses and couplings by
diagonalizing the $5 \times 5$ Higgs boson mass matrix, which was calculated
taking into account radiative corrections up to the dominant two loop terms, as
described in the previous section. The five mass eigenstates are denoted by
$h_i, i=1..5$ with masses $m_i$ in increasing order. We also computed the
charged Higgs boson mass, the stop and the chargino masses, and applied all the
available experimental constraints on these particles from LEP
\cite{LEP,LEPHA,LEPCH,LEPCHARGI,LEPSTOP}, as discussed below.

\begin{figure}
\begin{center}
\includegraphics[height=7cm]{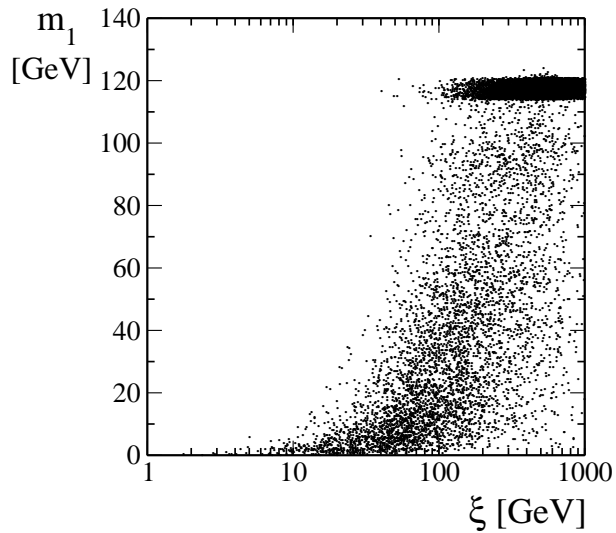}
\caption{Mass of the lightest Higgs state as a function of $\xi$, for a
 $m_t^{\rm pole} = 175 \gev$, $M_{\rm SUSY} = 350 \gev$ and maximal stop
 mixing. The other parameters are randomly chosen as in Eq.~(\ref{param}).}
 \label{fig1}
\end{center}
\end{figure}

The first important result is that we have succeeded in obtaining a large
number of points that complied with all the imposed constraints for {\it any
values of the CP violating phases $\t$ and $\d$}, \ie it is possible to have
SCPV in the NMSSM. For this result to be valid, the presence of the extra
tadpole term $\xi$ is crucial. This is easily understood from the fact that in
the limit where $\xi$ goes to zero, SCPV is no longer possible at tree level
\cite{Jorge}, and although viable when radiative corrections are included, the
Georgi-Pais theorem \cite{GP} predicts the appearance of light Higgs states in
the spectrum, already excluded by LEP. However, when $\xi \neq 0$, CP can be
spontaneously broken at tree level, as noted in the previous section. In
Fig.~\ref{fig1} we display the mass of the lightest Higgs boson, $m_1$, as a
function of the tadpole parameter $\xi$, with the other parameters randomly
chosen as in Eq.~(\ref{param}), for a set of approximately $10^4$ points. We
can see that small values of $\xi$ are associated with a very light mass for
the $h_1$ state. Such light Higgs states are not excluded by current
experimental bounds as long as their reduced coupling to the gauge bosons is
small enough. Indeed, the LEP exclusion limits on Higgs boson production gives
an upper bound on the reduced coupling of a Higgs boson to the gauge bosons as
a function of its mass. The reduced coupling $R_i$ is defined as the coupling
$ZZh_i$, divided by the corresponding Standard Model coupling:

\begin{equation}
R_i = h_{i1} \cos\b + h_{i2} \sin\b \; ,
\end{equation}

\noi where $h_{i1}, h_{i2}$ are the $S_1, S_2$ components of the Higgs state
$h_i$, respectively. One has $0 < |R_1| < 1$ and unitarity implies

\begin{equation}
\sum_{i=1}^5 R_i^2 = 1 \; .
\end{equation}

\begin{figure}
\begin{center}
\includegraphics[height=7cm]{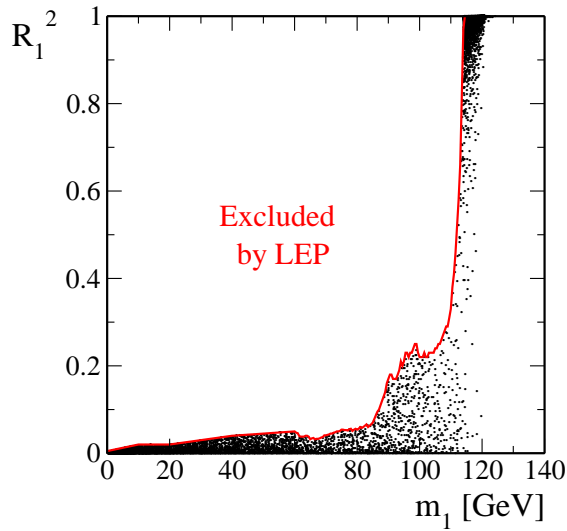}
\caption{Reduced coupling of the lightest Higgs state to the SM gauge bosons as
 a function of its mass for the same points as in Fig.~\ref{fig1}. The 
 solid line indicates the LEP exclusion limit \cite{LEP}.} \label{fig2}
\end{center}
\end{figure}

In Fig.~\ref{fig2} we plot the square of the reduced coupling of the lightest
Higgs state, $R_1^2$, versus $m_1$ for the same set of points as in
Fig.~\ref{fig1}. We also display the LEP exclusion curve \cite{LEP}, from which
one can see that the light $h_1$ states are indeed not excluded. The presence
of a light Higgs state with a small reduced coupling to the SM gauge bosons
might prove difficult to detect at future colliders. However, it is worth
stressing that in our results we always have at least one light {\em visible}
Higgs state with a large reduced coupling to the gauge bosons ($R_i^2 \gsim
0.5$) and a mass in the interval $112 \gev \lsim m_i \lsim 150 \gev$. The lower
bound is fixed by the LEP limit as shown in Fig.~\ref{fig2}. The upper bound
can be understood from the following relation

\begin{equation} \label{upperb}
\sum_{i=1}^5 R_i^2 m_i^2 = \cos^2\b M_{11}^2 + 2\cos\b\sin\b M_{12}^2 +
\sin^2\b M_{22}^2 \; ,
\end{equation}

\noi where $M_{11}^2, M_{12}^2, M_{22}^2$ are the Higgs boson mass matrix
elements as given in the previous section. The right hand side of
Eq.~(\ref{upperb}) gives the usual NMSSM upper bound on the lightest Higgs
boson mass \cite{CH}, with the only difference being that the stop mixing
parameter $X_t$  now depends on the CP violating phase $\t$, as seen in
Eq.~(\ref{Xtdef}). 

Assuming $m_t^{\mathrm pole} = 175 \gev$, $M_{\mathrm SUSY} = 350 \gev$ and
maximal stop mixing, one obtains $m_1 \lsim 125 \gev$ with this upper bound
being saturated for small $\tan\b$, namely $\tan\b \simeq 2.7$ \cite{CH}. One
can check from Figs.~\ref{fig1} and \ref{fig2} that this upper bound is only
reached by a few points in our set. The dense band of points in Fig.~\ref{fig1}
corresponds to cases where $h_1$ is a SM-like Higgs state with $R_1 \simeq 1$.
In this case, the lower bound from LEP is $m_1 \gsim 114 \gev$ and the upper
bound, from Eq.~(\ref{upperb}), is $m_1 \lsim 125 \gev$ as explained above.

We have also applied the exclusion limit from LEP on Higgs bosons associated
production ($e^+ e^- \to hA$ in the MSSM). This provides an upper bound on the
$Zh_ih_j$ reduced coupling as a function of $m_i+m_j$ for $m_i\simeq m_j$
\cite{LEPHA}. The $Zh_ih_j$ reduced coupling is the equivalent of $\cos(\b-\a)$
in the MSSM and is here defined as

\begin{equation}
R'_{ij} = h_{i4} (h_{j2} \cos\b - h_{j1} \sin\b) - ( i \leftrightarrow j ) \; ,
\end{equation}

\noi with $h_{i4}$ the CP-odd doublet $P$ component of $h_i$. We have also
taken into account the LEP limit on the charged Higgs boson mass \cite{LEPCH},
which here simply reads $m_{H^\pm} > 89.6 \gev$. By inspection of
Eq.~(\ref{chhmass}), this limit translates into having opposite signs for the
phases $\t$ and $\d$.

As we will see in the next section, charginos play an important role in the
computation of $\eK$. The tree level chargino mass matrix in the $(\wino,
\higgsino)$ basis reads

\begin{equation}
{\cal M}_{\charg} = \left( \ba{cc} M_2 & \sqrt{2} M_W \sin\b e^{-i\t} \\
\sqrt{2} M_W \cos\b & - \l x \ea \right) \; ,
\end{equation}

\noi where $M_2$ is the soft wino mass. We randomly scanned in the following
interval for $M_2$

\begin{equation} \label{M2range}
100 \gev < M_2 <250 \gev
\end{equation}

\noi and applied the LEP bound on the chargino mass, $m_{\chargi{1,2}} > 103.5
\gev$ \cite{LEPCHARGI}. Finally, we checked that for each point the stop masses
also satisfied the associated LEP limit $m_{\st{1,2}} > 100 \gev$
\cite{LEPSTOP}.

\mysection{Indirect CP violation in $K^0-\bar{K}^0$ mixing} \label{sec:epsK}

In the framework of the NMSSM with SCPV, all the SUSY parameters are real. Even
so, the physical phases of the Higgs doublets and singlet appear in the scalar
fermion, chargino and neutralino mass matrices, as well as in several
interaction vertices. In this section we will explore whether or not these
physical phases can account for the experimental value of $\eK$,  $\eK=(2.271
\pm 0.017) \times 10^{-3}$~\cite{Hagiwara}.

It is worth stressing that in this scenario the SM does not provide any
contribution to the CP violating observables, since the CKM matrix is real.
This can be clarified by noting that since $N$ is a singlet field, it does not
couple to the quarks, and although both Higgs doublets do couple, the phase
associated with these couplings can be rotated away by means of a redefinition
of the right-handed quark fields. Since charged currents are purely
left-handed, these phases do not show up in the $\vckm$, which is thus real.

Let us now proceed to compute the contributions to the indirect CP violation
parameter of the kaon sector, which is defined as

\begin{equation} \label{eK:eq:def}
\eK \simeq \frac{e^{i\pi/4}}{\sqrt{2}} \frac{\operatorname{Im}
\mathcal{M}_{12}}{\Delta m_K} \; .
\end{equation}

\noi In the latter $\Delta m_K$ is the long- and short-lived kaon mass
difference, and $\mathcal{M}_{12}$ is the off-diagonal element of the neutral
kaon mass matrix, which is related to the effective hamiltonian that governs
$\Delta S=2$ transitions as

\begin{equation} \label{eK:M12:def}
\mathcal{M}_{12} = \frac{\vev{K^0|\mathcal{H}_{\text{eff}}^{\Delta S=2}
|\bar{K}^0}}{2m_K} \; , \quad \text{with} \qquad
\mathcal{H}_{\text{eff}}^{\Delta S=2} = \sum_i c_i {\mathcal O}_i \; .
\end{equation}

\begin{figure}
\begin{center}
\includegraphics[height=7cm]{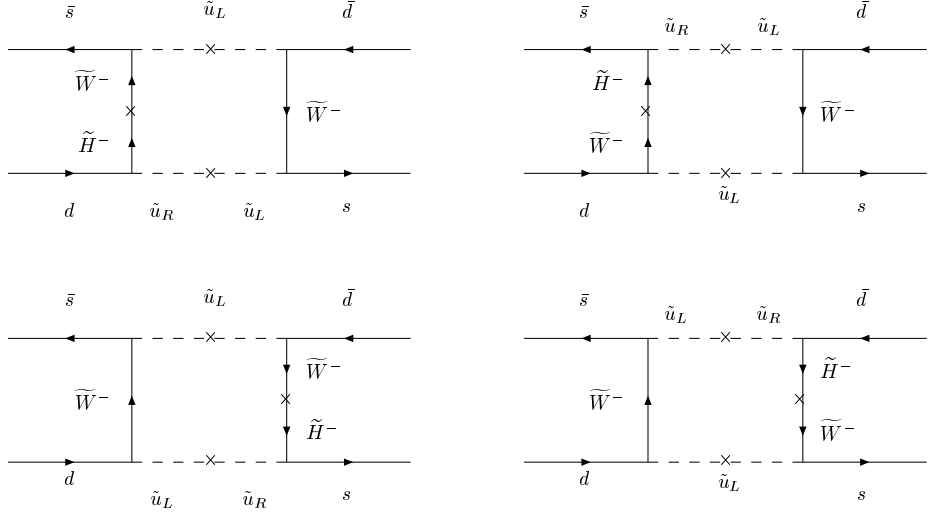}
\caption{Box diagrams associated with the leading chargino contribution to
 $\eK$.} \label{fig:eK:4box}
\end{center}
\end{figure}

\noi Here $c_i$ are the Wilson coefficients and ${\mathcal O}_i$ the
local operators. In the presence of SUSY contributions, the Wilson
coefficients can be decomposed as $c_i= c_i^W + c_i^{H^\pm}+
c_i^{\tilde{\chi}^\pm} + c_i^{\tilde{g}} + c_i^{\tilde{\chi}^0}$. As
discussed in~\cite{Lisbon}, in the present class of models where there
are no contributions from the SM, the chargino mediated box diagrams
give the leading supersymmetric contribution, and the $\Delta S=2$
transition is largely dominated by the $(V-A)$ four fermion operator
$\mathcal{O}_1$. Working in the weak basis for the
$\tilde{W}-\tilde{H}$, rather than in the physical chargino basis, and
using the mass insertion approximation for the internal squarks, we
have verified that the $\eK$ receives the leading contribution from
the box diagrams depicted in Fig.~\ref{fig:eK:4box}. In the limit of
degenerate masses for the left-handed up-squarks, $\operatorname{Im}
\mathcal{M}_{12}$ is given by~\cite{Lisbon}

\begin{eqnarray} \label{eK:M12:equation}
\operatorname{Im} \mathcal{M}_{12} & = & \frac{2 G_F^2 f^2_K m_K m_W^4
}{3\pi^2\vev{m_{\tilde{q}}}^8} (V_{td}^*V_{ts}) m_t^2 \left| e^{i\t} \;
m_{\tilde{W}} - \cot\beta \; m_{\tilde{H}} \right| \nn \\
& & \times \left\{ \Delta A_U \sin[\varphi_{\chi}-\t] \;
{(M^2_{\tilde{Q}})}_{12} \; I(r_{\tilde{W}}, r_{\tilde{H}}, r_{\tilde{u}_L},
r_{\tilde{t}_R}) \right\} \; .
\end{eqnarray}

\noi In the above equation, $f_K$ is the kaon decay constant and $m_K$ the kaon
mass~\cite{Hagiwara}; $V_{ij}$ are the $\vckm$ elements, whose numerical values
($V_{td}=0.0066$ and $V_{ts}=-0.04$) reflect the fact that we are dealing with
a flat unitarity triangle; $\vev{m_{\tilde{q}}}$ is the average squark mass,
which we take equal to $M_{\rm SUSY}$; $m_{\tilde{W}} = M_2$ is the wino mass,
$m_{\tilde{H}} = \mu$ is the higgsino mass and $\varphi_{\chi} = \arg (
e^{i\theta}m_{{\wino}} - \cot\beta\; m_{{\higgsino}} )$\footnote{Our
conventions differ from Ref.~\cite{Lisbon} in that $\mu_{eff} \rightarrow
-\mu$.}. The non-universality in the $LL$ soft breaking masses is parametrized
by ${(M^2_{\tilde{Q}})}_{12}$, which we choose taking into account the bounds
from the analysis in Ref.~\cite{MQ12}. As discussed in Ref.~\cite{Lisbon}, a
sizable non-universality in the soft trilinear terms is crucial, and is here
parametrized by $\Delta A_U\equiv A_U^{13}-A_U^{23}$. In the following, we
assume $\D A_U = 500 \gev$. Finally, $I$ is the loop function, with
$r_i=m_i^2/\vev{m_{\tilde{q}}}^2$~\cite{Lisbon}.

By scanning the parameter space of the model, we have verified that one can
find sets of parameters that satisfy the minimization conditions of the Higgs
potential, have an associated Higgs boson spectrum compatible with LEP searches
and still succeed in generating the observed value of $\eK$. From
Eq.~(\ref{eK:M12:equation}), it appears that $\eK$ depends on $1/M_{\rm
SUSY}^8$. It is therefore difficult to saturate the experimental value of $\eK$
for large values of $M_{\rm SUSY}$. On the other hand, too small values of
$M_{\rm SUSY}$ might generate a light Higgs boson spectrum, already excluded by
LEP. In order to accommodate both constraints, we took $M_{\rm SUSY} = 350
\gev$, as already referred to in section~\ref{sec:higgs}. All other parameters
are as in Eqs.~(\ref{param}, \ref{M2range}) and we assumed $m_t^{\rm pole} =
175 \gev$ and maximal stop mixing, as in the previous section.

\begin{figure}
\begin{center}
\includegraphics[height=65mm]{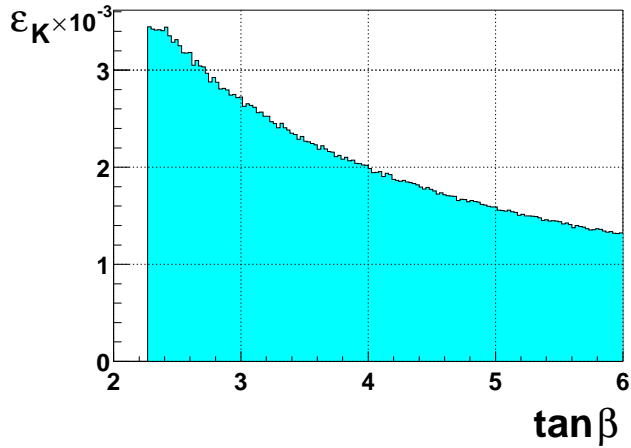}
\caption{Values of $\eK$ as function of $\tan\beta$ (shaded area) for $m_t^{\rm
 pole} = 175 \gev$, $M_{\rm SUSY} = 350 \gev$ and maximal stop mixing. The other
 parameters are randomly chosen as in Eqs.~(\ref{param}, \ref{M2range}).}
\label{tbeK}
\end{center}
\end{figure}

In Fig.~\ref{tbeK}, we plot  the possible values of $\eK$ as a function of
$\tan\b$. The maximal values of $\eK$ are obtained for the low $\tan\beta$
regime, and the experimental bounds on $\eK$ require $\tan\b \lsim 3.8$. Recall
that the lower bound for $\tan\beta$ is consistent with the analysis conducted
in section~\ref{sec:higgs}. Apart from the explicit dependence of
Eq.~(\ref{eK:M12:equation}) on $\tan\beta$, one should bear in mind that there
are also implicit dependences associated with the other parameters  as well as
with the various experimental bounds imposed on the mass spectrum. It is
therefore difficult to reproduce analytically the observed upper bound on $\eK$
as a function of $\tan\b$.

\begin{figure}
\begin{center}
\includegraphics[height=65mm]{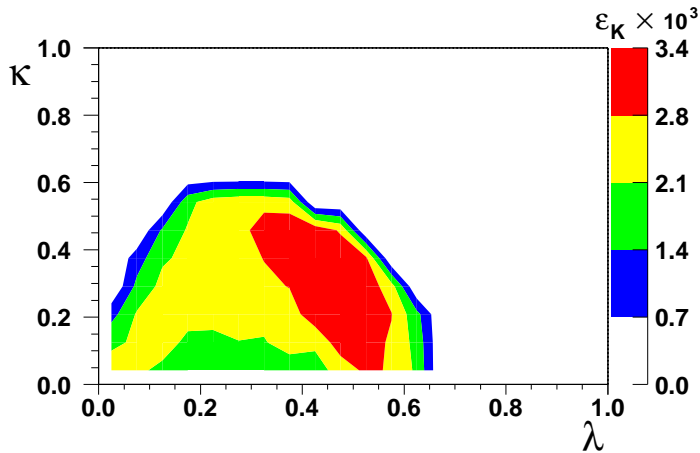}
\caption{Contours for the maximum value of $\eK$ in the $\lambda$--$\kappa$
 plane for $m_t^{\rm pole} = 175 \gev$, $M_{\rm SUSY} = 350 \gev$, maximal stop
 mixing and the other parameters as in Eqs.~(\ref{param}, \ref{M2range}).}
\label{lkcontours}
\end{center}
\end{figure}

\begin{figure}
\begin{center}
\includegraphics[height=65mm]{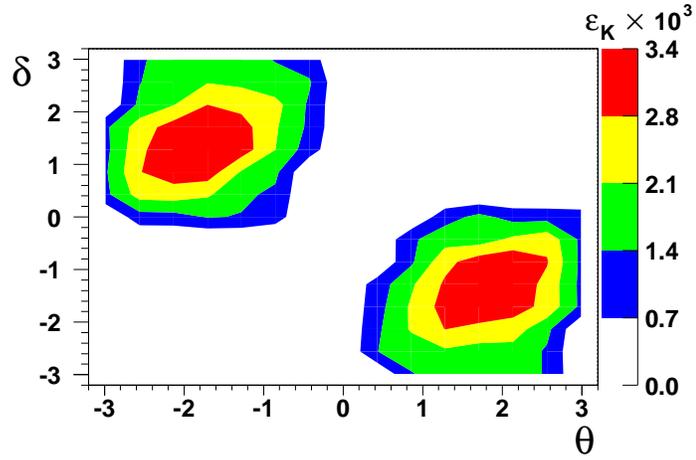}
\caption{Contours for the maximum value of $\eK$ in the $\theta$--$\delta$
 plane. All parameters as in Fig.~\ref{lkcontours}.}
\label{tdcontours}
\end{center}
\end{figure}

\begin{figure}
\begin{center}
\includegraphics[height=65mm]{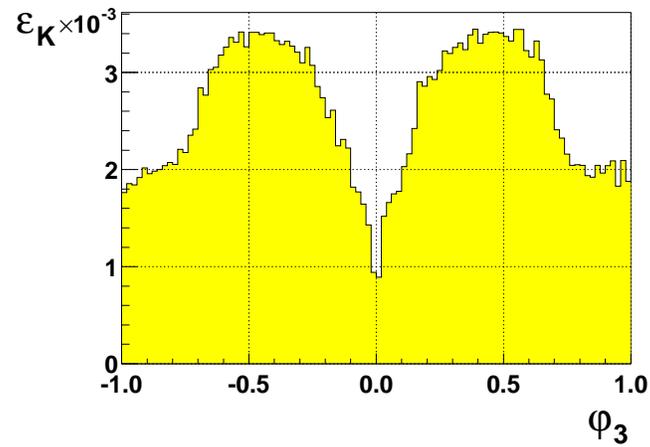}
\caption{Values of $\eK$ as function of the singlet phase
 $\varphi_3=\d/3$ (grey area). All parameters as in Fig.~\ref{lkcontours}.}
\label{PNeK}
\end{center}
\end{figure}

In Fig.~\ref{lkcontours}, we plot the maximal value of $\varepsilon_K$ in
distinct regions of the $\lambda - \kappa$ plane. The remaining parameters are
chosen in order to maximize $\varepsilon_K$ and still comply with the
experimental bounds. As one can see from this figure, having $\varepsilon_K
\sim 2\times 10^{-3}$ is associated with values of $\kappa$ and $\lambda$ in
the range $[0, 0.6]$. In other words, one can easily saturate $\varepsilon_K$
in a vast region of the parameter space.

As expected from the inspection of Eq.~(\ref{eK:M12:equation}), there is a
strong dependence of $\eK$ on the phases associated with the Higgs fields. In
Fig.~\ref{tdcontours}, we display contour plots for the maximal values of $\eK$
in the plane generated by the phases $\t$ and $\d$. Although,  {\it a priori},
all the values for the phases $\t$ and $\d$ in $[-\pi,\pi]$ are allowed, it is
clear from Fig.~\ref{tdcontours} that the saturation of the experimental value
of $\eK$ can only be achieved for significant values of the singlet and doublet
phases.  This feature becomes more evident in Fig.~\ref{PNeK}, where we show
the values of $\eK$ as a function of the singlet phase, $\varphi_3 = \delta/3$.
The saturation of the experimental value of $\eK$ requires the singlet phase to
be $|\varphi_3| \gtrsim 0.15$. We will discuss in the conclusions the
implications of such large CP violating phases.

\mysection{Conclusions and discussion}

We have shown that in the framework of the NMSSM with an extra tadpole term for
the singlet in the scalar potential, it is possible to have spontaneous CP
violation while complying with the LEP constraints on the Higgs boson mass.
Although in this scenario $\vckm$ is real, the experimental value of $\eK$ can
be saturated in large parts of the parameter space of the model, without
requiring any fine tuning. This may be of special interest for electroweak
baryogenesis, as it has been shown that a strong first order electroweak phase
transition is possible in the NMSSM, as opposed to the MSSM, due to the
additional trilinear Higgs boson couplings in the lagrangian \cite{baryo}.

The presence of the extra tadpole term $\xi$ is crucial to conciliate
spontaneous CP violation with the LEP constraints on the Higgs boson mass.
Moreover, the $\xi$ term also solves the domain wall problem associated with
the spontaneous breaking of the $\mathbb{Z}_3$ symmetry, which is present if
$\xi = 0$. On the other hand, since CP is spontaneously broken at the
electroweak scale, CP domain walls may appear, which are cosmologically
excluded. An elegant solution to this problem, along the lines of the
$\mathbb{Z}_3$ domain walls solution, is to assume that gravitational
interactions also explicitly break CP. In the low energy scalar potential, one
then obtains an explicit CP violating tadpole term, \ie a complex $\xi$ term. A
phase in the $\xi$ term would not change the results derived here, since its
only effect is to generate a shift in the singlet phase $\d$.

Regarding the large phase regime favoured by the saturation of $\eK$, one
should recall that $\t$ and $\d$ are flavour-conserving phases, and might
generate sizable contributions to the electron, neutron and mercury atom EDMs.
Although we will not address the EDM problem here, a few remarks are in order:
first, let us notice that in the presence of a small singlet coupling $\l$, as
allowed in our results (see Fig.~\ref{lkcontours}), the EDM constraints on $\d$
become less stringent~\cite{tanimoto}. In addition, there are several possible
ways to evade the EDM problem, namely reinforcing the non-universality on the
trilinear terms (\ie requiring the diagonal terms to be much smaller than the
off-diagonal ones or having matrix-factorizable $A$ terms), the existence of
cancellations between the several SUSY contributions, and the suppression of
the EDMs by a heavy SUSY spectrum~\cite{Khalil}. In view of the considerably
large parameter space allowed in our results, none of these possibilities
should be disregarded.

Concerning the other CP-violating observables, namely  $\varepsilon^\prime /
\varepsilon$ and the CP asymmetry of the $B_d$ meson decay ($a_{J/\psi K_S}$),
it has been pointed out that this class of models can generate sizable
contributions, although saturating the experimental values generally favours a
regime of large phases and maximal $LR$ squark mixing \cite{Pomarol2,Lebedev}.
A complete analysis of these issues, as well as of the EDM problem in our model
has yet to be done.

In this work, we pointed out that  the NMSSM with an extra tadpole term in the
scalar potential appears to be an excellent candidate for a scenario of SCPV. 
Having a $\mathbb{Z}_3$ invariant superpotential preserves the original
motivation of the NMSSM to solve the $\mu$ problem of supersymmetry. The extra
tadpole term for the singlet cures the domain wall problem of the
$\mathbb{Z}_3$ invariant model and allows the spontaneous breaking of CP. In
this model, one can simultaneously saturate $\eK$ and obtain a particle 
spectrum compatible with the current experimental bounds.

The presently available experimental data on CP violation does not allow to
distinguish whether the CP symmetry is spontaneously or explicitly broken. 
With the advent of the LHC and eventually of the Linear Collider, direct
searches of the Higgs bosons, and the measurement the associated couplings to
the SM gauge bosons, will allow to disentangle this model from the MSSM.

\pagebreak
\noi {\bf \Large Acknowledgments}
\vskip .5 truecm

\noi This work was supported by the Spanish grant BFM2002-00345 and by
the European Commission RTN grant HPRN-CT-2000-00148. A.~T. was
supported by {\it Funda\c c\~ao para a Ci\^encia e Tecnologia} under
the grant SFRH/BPD/11509/2002. J.C.R. was partly supported by the
Marie Curie fellowship HPMF-CT-2002-01902.

\end{document}